\documentclass[fleqn,twoside,twocolumn,nofootinbib,showkeys,aps]{revtex4}

\usepackage[nocpr]{ujp}

\begin{document}

\title[Elements of $\mu$-calculus and thermodynamics of $\mu$-Bose gas]
{ELEMENTS OF \boldmath$\mu$-CALCULUS\\ AND THERMODYNAMICS OF \boldmath$\mu$-BOSE GAS MODEL$^1$}
\author{A.P.~Rebesh}
\affiliation{Bogolyubov Institute for Theoretical Physics, Nat. Acad. of Sci. of Ukraine}
\address{14b, Metrolohichna Str.
 Kyiv 03680, Ukraine}
\email{rebesh@bitp.kiev.ua, omgavr@bitp.kiev.ua}
\author{I.I.~Kachurik}
\affiliation{Bogolyubov Institute for Theoretical Physics, Nat. Acad. of Sci. of Ukraine}
\address{14b, Metrolohichna Str., Kyiv 03143, Ukraine}
\affiliation{Khmelnytskyi National University}
\address{11, Instytutska Str., Khmelnytskyi 29016, Ukraine}
\author{A.M.~Gavrilik}
\affiliation{Bogolyubov Institute for Theoretical Physics, Nat. Acad. of Sci. of Ukraine}
\address{14b, Metrolohichna Str.
 Kyiv 03680, Ukraine}
\email{omgavr@bitp.kiev.ua}

\udk{536}  \razd{\secix}

\autorcol{A.P.\hspace*{0.7mm}Rebesh, I.I.\hspace*{0.7mm}Kachurik,
A.M.\hspace*{0.7mm}Gavrilik}

\setcounter{page}{1182}

\begin{abstract}
    We review on and give some further details about the thermodynamical properties of the
$\mu$-Bose gas model (arXiv:1309.1363) introduced by us recently. This model was elaborated in connection with
$\mu$-deformed oscillators.
Here, we present the necessary concepts and tools from the so-called $\mu$-calculus. For the high temperatures,
we obtain the virial expansion of the equation of state, as well as five virial
coefficients. In the regime of low temperatures, the critical temperature of condensation is inferred. We also
obtain the specific heat, internal energy, and entropy for a $\mu$-Bose gas for both low and high temperatures.
All thermodynamical functions depend on the deformation parameter $\mu$.
The dependences of the entropy and the specific heat on the deformation parameter are visualized.
\end{abstract}


\maketitle
    \section{Introduction}\vspace*{2mm}
Quantum algebras and deformed oscillator algebras play an important
role in the description of complex systems in different fields of
physics, especially of systems with essential nonlinearities.

In modern quantum  field theory, the use of deformed oscillators is
motivated, in particular, by the necessity to incorporate the exotic
statistics of particles, e.g., that of quons \cite{Greenberg}. On
the other hand, in the two-dimensional case, the anyonic \cite{Wilczek}
fractional statistics (connected with the braid group) requires
modified commutation relations and is used in the description of
the fractional quantum Hall effect and high-temperature
superconductivity \cite{ChenWWH}. Moreover, the discrepancy between
theory and experiments that concerns the unstable phonon spectrum
in ${}^4$He can be nicely overcome if one treats phonons, by using
bosonic deformed oscillators \cite{Monteiro}. The application of
deformed oscillator algebras is efficient in the phenomenological
study of the properties of elementary particles produced in
relativistic heavy ion collisions. Specific deformed oscillator
models, as the base for a respective deformed Bose gas model, are
applied to the issue of (the intercepts of) correlation functions of
hadrons measured in the experiments on heavy-ion collisions
(STAR/RHIC) and turn out to be successful
\mbox{\cite{AGP,GavrS,GavrRebIntercepts}.}\looseness=1

\footnotetext[1]{This work is the contribution to Proceedings of the
International Conference ``Quantum Groups and Quantum Integrable
Systems''.}Generally speaking,  the attempts to modify (deform) the
commutation relations of an ordinary quantum harmonic oscillator were
encountered many times in theoretical physics \cite{Kobayashi}.
Among different deformed oscillator models, the simplest and most
popular are the Arik--Coon \cite{ArikCoon76} and Biedenharn--Macfarlane
\cite{Biedenharn,Macfarlane89} models and a two-parameter
$p,q$-deformed oscillator \cite{Chakrabarti91}. The basic properties
and diverse applications of these three types of deformed
oscillators are intensively studied. In addition, the Tamm--Dancoff
deformed oscillator model and the Jannussis  $\mu$-oscillator
\cite{Jann} attract more and more attention during \mbox{last
years.}

Basing on a set of deformed oscillators, diverse deformed analogs
of the Bose gas model can be developed. The thermodynamics of $q$-Bose
and $p,q$-Bose gases have been investigated at high and low
temperatures \cite{ScarfoneSwamy08,ScarfoneSwamy09,AlginFibOsc}. In
this contribution that follows \cite{Arxiv}, we explore
the thermodynamics of another deformed analog of the Bose gas
model ($\mu$-Bose gas) recently elaborated by us in
\cite{GavrRebIntercepts} in more details. Namely, we deal with the system of
$\mu$-deformed oscillators (gas of $\mu$-bosons). Here, $\mu$ is the
deformation parameter.

On the other hand, to study the thermodynamics of such system, one could
start with a deformed Hamiltonian. But there is a more efficient
and significantly simpler way to obtain thermodynamical functions
and values that uses some specially designed analog of calculus
(here, $\mu$-calculus).

An example of generalized calculus including the $q$-derivative and
the $q$-integral was proposed long ago by Jackson \cite{Jackson}. This
$q$-calculus arises in the description of systems with discrete
dilatation symmetries \cite{Erzan}. Of course, it appears as a
convenient tool for the study of the thermodynamics of a $q$-Bose gas.
Somewhat later, the $q$-calculus was generalized to the $p,q$-calculus.

Inspired by the successes of earlier known applications of
the $q$-calculus, we introduce the necessary elements
of the so-called $\mu$-calculus. Its usage allows us to overcome
encountered difficulties in obtaining the thermodynamical functions of
the $\mu$-Bose gas model.

The paper is organized  as follow. In Section 2, we introduce the
elements of the $\mu$-calculus that will be necessary in the study of
the thermodynamics of a $\mu$-Bose gas. The deformed analogs of some
elementary functions (exponential, logarithmic, {\it etc}.) are also given as
examples. Sections 4 and 5 are devoted to the deformed analog of
the Bose gas model. We introduce three different approaches for
a deformation of a Bose gas and then focus on the particular $\mu$-Bose
gas model. In Section 5, we obtain the total number of particles,
virial coefficients of the equation of state, critical temperature,
and some other thermodynamical functions.

\section{Deformed Oscillators\\ and Deformations of Calculus}
We  start with the algebra of a deformed oscillator, which
can be given, in general, by the following commutation relation:
\begin{equation}
[a,a^{\dagger}]=\varphi(N)-\varphi(N+1), \quad
aa^{\dagger}\equiv\varphi(N),
\end{equation}
where $\varphi(N)$  is the structure function \cite{Polynomially},
which determines a specific model of deformed oscillator. The
commutation relations for the creation and annihilation operators
$a$, $a^{\dagger}$ and the operator of particle number read
\begin{equation}
[a,a]=[a^{\dagger},a^{\dagger}]=0,\quad
[N,a^{\dagger}]=a^{\dagger},\quad [N,a]=-a.\!\!\!
\end{equation}
The transformation from the Fock space to the configuration space
(Bargmann-like representation) can be done by the replacement
\begin{equation}
a^{\dagger} \rightarrow x, \qquad a \rightarrow \mathcal{D}_x^{\rm
(def)},
\end{equation}
where $\mathcal{D}_x^{\rm (def)}$ is some analog of the usual
derivative (say, that introduced by Jackson).

\subsection*{Jackson derivative and its
extensions}

To derive the thermodynamical functions for a deformed analog of the Bose gas model, it is necessary to modify the
usual treatment. First of all, this concerns the derivatives (calculus) \cite{Kac}. In particular,
if the Jackson
or $q$-derivative
 \begin{equation}\label{4}
\bigg(\!\frac{d}{dx}\!\bigg)_{\!\!q}f(x)\equiv  \mathcal{D}^{(q)}_x
f(x)=\frac{f(qx)-f(x)}{qx-x}
\end{equation}
instead of the usual derivative is in use, this leads to a consistent generalization.
Taking the limit $q\rightarrow 1$ recovers $d/dx$, i.e.,
\begin{equation}\label{5}
\mathcal{D}_x^{(q)}\stackrel{q\rightarrow1}{\longrightarrow}\frac{d}{dx}.
\end{equation}
On the monomial $x^n,$ the $q$-derivative gives
\begin{equation}\label{6}
\mathcal{D}^{(q)}_x x^n\!=\!\frac{(qx)^n-x^n}{qx-x}\!=\![n]_qx^{n-1}, \quad [n]_q\!\equiv\!\frac{q^n-1}{q-1},
\end{equation}
where $[n]_q$\! denotes the $q$-bracket. As $q\!\!\rightarrow\! 1$,
$\displaystyle\lim_{q\rightarrow1}[n]_q\!\!=$ $=n,$ and the action
of the ordinary derivative is recovered. Under the action of
$x\mathcal{D}^{(q)}_x$, the monomials behave as eigenvectors. It is
easy to check that now the ``coordinate'' and ``momentum'' operators
realized by $X\rightarrow x$ and $P\rightarrow \mathcal{D}^{(q)}_x$
lead to the algebra of an Arik--Coon-type deformed oscillator with
$aa^{\dagger}-qa^{\dagger}a=1,$ where $q$ is the deformation
parameter.

A two-parameter $p,q$-extension of the Jackson derivative is also known \cite{Floreanini93,BurbanKlimyk}.
This is the $p,q$-derivative $\mathcal{D}^{(p,q)}_x,$ which acts as
\begin{equation*}\label{6+1}
\mathcal{D}^{(p,q)}_xf(x)=\frac{f(px)-f(qx)}{px-qx},
\end{equation*}\vspace*{-7mm}
\[\mathcal{D}^{(p,q)}_xx^n=[n]_{p,q}x^{n-1}, \quad
[n]_{p,q}\equiv\frac{p^n-q^n}{p-q}.
\]
When $p=1,$ we pass back to the Jackson derivative $\mathcal{D}^{(q)}_x $ given in (\ref{4}) and (\ref{6}).

Note that a slightly different $p,q$-extension $\widetilde{\mathcal{D}}_z^{(p,q)}$
of the Jackson derivative such that
\begin{equation*}\label{a}
\widetilde{\mathcal{D}}_z^{(p,q)}x^k=\frac{p^k-q^k}{\ln p-\ln q}x^{k-1}=\frac{p-q}{\ln p-\ln q}\mathcal{D}^{(p,q)}_xx^k
\end{equation*}
was adopted in \cite{AlginFibOsc}.

\subsection*{Elements of \boldmath$\mu$-calculus}
As some alternative to the Jackson derivative, we will use the ``$\mu$-derivative''. It is important
for the construction of the $\mu$-Bose gas model. Namely, we introduce the $\mu$-extension of $d/dx$ or the $\mu$-analog of
the derivative ($\mu$-derivative) acting on $x^n$ as follows:
\begin{equation}\label{7}
\mathcal{D}^{(\mu)}_x x^n=[n]_{\mu}x^{n-1}, \quad [n]_{\mu}\equiv\frac{n}{1+\mu n}.
\end{equation}
As $\mu \rightarrow 0,$ the usual $d/dx$ is recovered from $\mathcal{D}^{(\mu)}_x$.
This special form of a deformed derivative is inspired, due to the appearance of the
$\mu$-bracket, by a Jannussis $\mu$-deformed oscillator \cite{Jann}. As $\mu\rightarrow 0,$ the $\mu$-extension
$\mathcal{D}^{(\mu)}_x$ goes over directly into the usual derivative.

Though the formula showing how the $\mu$-derivative acts on monomials $x^m$ is sufficient for our goals,
let us also indicate how the $\mu$-derivative does operate upon a generic function $f(x)$:
\begin{equation}\label{7-a}
\mathcal{D}^{(\mu)}_xf(x)\!=\!\int\limits^1_0dtf'_x(t^{\mu}x), \quad
f'_x(t^{\mu}x)\!=\!\frac{d f(t^{\mu}x)}{d x}.
\end{equation}
As easily seen, the above formula (\ref{7}) stems from this general definition.

The $k$-th power of the $\mu$-derivative acting on the monomial $x^n$ yields
\begin{equation}\nonumber
(\mathcal{D}^{(\mu)}_x)^kx^n=\frac{[n]_{\mu}!}{[n-k]_{\mu}!}x^{n-k},
\end{equation}
where
\begin{equation}\label{013}
[n]_{\mu}!\!=\!\frac{n!}{[n;\mu]}, \quad [n;\mu]\equiv[1+\mu][1+2\mu]...[1+n\mu].
\end{equation}
The extended versions of the $\mu$-derivative (\ref{7}), (\ref{7-a}), the $q,{\mu}$-
and $(p,q;\mu)$-derivatives can be also introduced: for this, we merely insert, respectively, $\mathcal{D}^{q}_xf(t^{\mu}x)$ and
$\mathcal{D}^{(p,q)}_xf(t^{\mu}x)$ in (\ref{7-a}) instead of $(d/dx)f(t^{\mu}x)$.
Such extensions naturally correspond to the $(q;\mu)$- and
$(p,q;\mu)$-deformed quasi-Fibonacci oscillators that appeared in \cite{Kachurik}.

The inverse $\mu$-derivative
$\big({\mathcal{D}^{(\mu)}_x}\big)^{-1}$ (or the anti-$\mu$-derivative)
for $D^{(\mu)}_x$ from (\ref{7}) and (\ref{7-a}) does also exist.

Below, dealing with a $\mu$-deformed analog of the Bose gas model, we will utilize, at a proper point, this finite-difference derivative $D^{(\mu)}_z$ instead of
the usual $d/dz$. In such a way, the deformation parameter gets imported in the model (of course, another root
for introducing a $\mu$-deformation is also possible).

\begin{figure*}
\vskip1mm
\includegraphics[width=16.7cm]{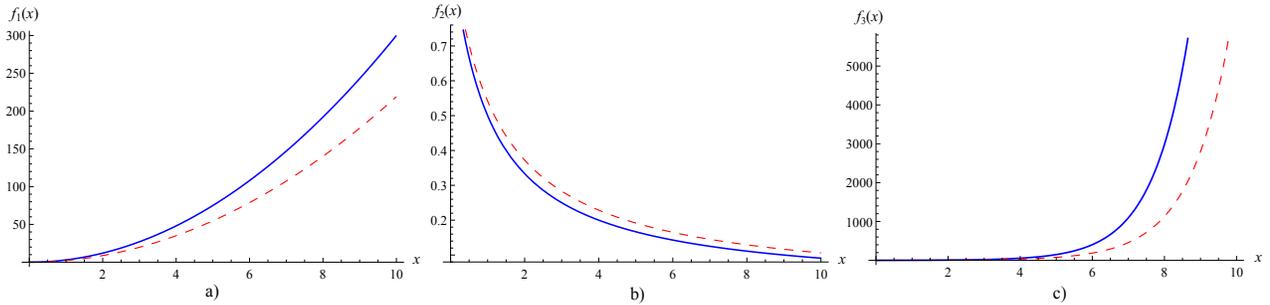}
\vskip-2mm\caption{Behavior of the usual derivative $d/dx$ (solid
line) with $\mu=0$ and its $\mu$-deformed analog
$\mathcal{D}^{(\mu)}_x$ (dashed line) with $\mu=0.7$ acting on
$f_1=x^{3}$, $f_2=\ln (1+x)$, and $f_3=\exp(x)$: ({\it a}),
({\it b}), and ({\it c}), respectively \label{Fig:1}}\vspace*{2mm}
\end{figure*}

It is worth to note that, for small  positive values of deformation
parameter, both the usual and deformed (special finite difference)
derivatives of a function have similar behavior. Figure \ref{Fig:1}
gives a comparison of the behavior of the usual derivative (blue
line) operating on the monomial an the logarithmic and exponential
functions with that of the $\mu$-analog of the derivative
$\mathcal{D}^{(\mu)}_x$ taken on the same functions. As seen, the
both extensions of the derivative show a similar behavior for small
values of deformation \mbox{parameter $\mu$.}

This fact may serve to partly justify\,\footnote[2]{Physical
consequences of a deformed model based on the $\mu$-cal\-culus may differ
from those of other deformed ones.} the use of the $\mu$-derivative,
instead of the ordinary one, in calculating some thermodynamical
quantities in the $\mu$-Bose gas model. Then, through the use of
the $\mu$-analog of the derivative, the deformation parameter gets involved
in the treatment, and, thus, the whole system is $\mu$-deformed.

It is important that, with the use of the $\mu$-bracket $[n]_{\mu}$ and
the $\mu$-factorial $[n]_{\mu}!$ given in  (\ref{7}) and (\ref{013}),
respectively, we can consider the $\mu$-deformed
analogs of elementary functions such as, e.g., the $\mu$-exponential function
$\exp_{\mu}(x)$,
\begin{equation*}\label{exp}
  \exp_{\mu}(x)\!=\!\sum_{n=0}^{\infty}\frac{x^n}{[n]_{\mu}!},
\end{equation*}
or the $\mu$-logarithmic function $\ln_{\mu}(x)$,
\begin{equation*}\label{ln}
  \ln_{\mu}(x)=-\sum_{n=1}^{\infty}\frac{(1-x)^n}{[n]_{\mu}}.
\end{equation*}
The $\mu$-number $[n]_{\mu}$ and the $\mu$-factorial $[n]_{\mu}!$
reduce, as $\mu\rightarrow 0,$ to  $n$ and $n!$, and we recover the
usual exponential and logarithmic functions.

\begin{figure}
\includegraphics[width=7.0cm]{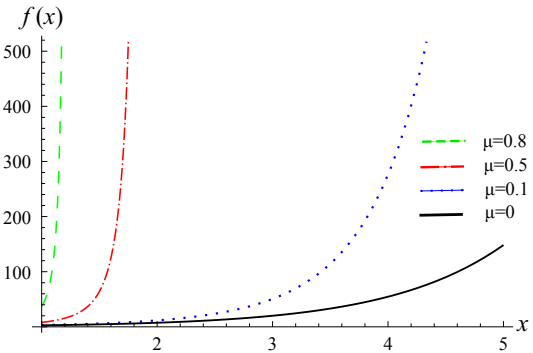}
\vskip-3mm\caption{$\mu$-Deformed analog of the exponential function:
$f(x)=\exp_{\mu}(x)$ \label{Fig-exp}}
\end{figure}

\begin{figure}
\vskip1mm
\includegraphics[width=7.0cm]{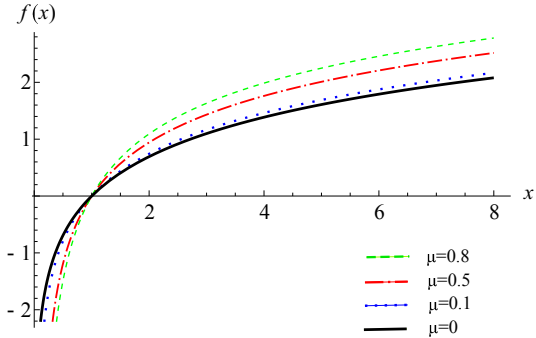}
\vskip-3mm\caption{$\mu$-Deformed analogs of the logarithmic function. Here,
$f(x)=\ln_{\mu}(x)$ \label{Fig-ln}}
\end{figure}

In Figs. \ref{Fig-exp} and \ref{Fig-ln},  the $\mu$-deformed analogs
of the exponential and logarithmic functions are shown. It is clear
that some special functions (e.g., the $\mu$-analog of
polylogarithms, see below) can also appear.

Let us introduce the $\mu$-analog ($\mu$-polylogarithm) of the
well-known Bose function or polylogarithm $g_l(z)=$
$=\sum_{n=1}^{\infty}z^n/n^l$, namely:\vspace*{-1mm}
\begin{equation}\label{16-14}
g_{l}^{(\mu)}(z)=\sum_{n=1}^{\infty}\frac{[n]_{\mu}}{n^{l+1}}z^n.
\end{equation}\vspace*{-3mm}

\begin{figure}[b!]
\vspace*{-3mm}
\includegraphics[width=7cm]{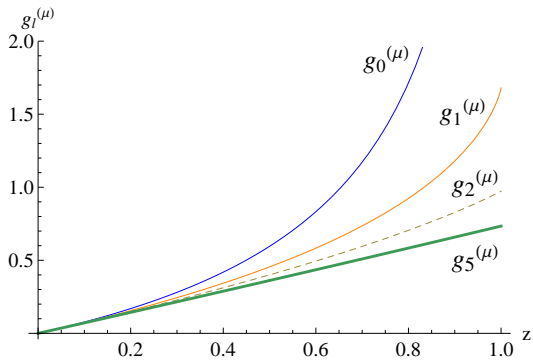}
\caption{Dependence of the functions $g_0^{\mu}$, $g_1^{\mu}$,
$g_2^{\mu}$, and $g_5^{\mu}$ on the fugacity
$z=\exp(\beta\widetilde{\mu})$ at $\mu=0.4$ \label{g}}\vskip1.5mm
\end{figure}

\noindent Its simplest special case is the function
$g_0^{(\mu)}(z)=$ $=\sum_{n=1}^{\infty}\frac{[n]_{\mu}}{n}z^n
\stackrel{\mu=0}{\longrightarrow}\sum_{n=1}^{\infty}z^n$. Clearly,
it reduces to $\ln(1-z)$ if $\mu=0$. In Fig.~\ref{g}, we illustrate
the dependence of the deformed Bose functions $g_0^{\mu}$,
$g_1^{\mu}$, $g_2^{\mu}$, and $g_5^{\mu}$ on the fugacity $z$ at
$\mu=0.4$.

\subsection*{\boldmath$\mu$-Derivative $D^{(\mu)}_x$\!\\ and $\mu$-analog of
the Leibnitz rule}

Consider how the $\mu$-differentiation acts upon the product $f(x)
g(x)$ of two functions. In the simple case of monomials $f(x)=x^n$
and $g(x)=x^m,$ we have\vspace*{-1mm}
\[
D^{(\mu)}_x \left( x^n x^m\right) = D^{(\mu)}_x \left( x^m
x^n\right) =
\]\vspace*{-7mm}
\begin{equation}\label{006}
=\frac{n+m}{1+\mu(n+m)}\, x^{n+m-1} .
\end{equation}
 More involved is the case of general functions,\vspace*{-1mm}
\begin{equation}\label{007}
D^{(\mu)}_x \left( f(x) g(x)\right) = \int\limits^1_0 [f(t^\mu x)
g(t^\mu x)]'_x {\rm d}t = \mbox{...}.
\end{equation}\vspace*{-3mm}

\noindent which After the integration by parts, we
obtain\vspace*{-1mm}
\[
 D^{(\mu)}_x \left( f(x) g(x)\right) = g(x) D^{(\mu)}_x f(x) + f(x) D^{(\mu)}_x g(x)-\]\vspace*{-5mm}
 \[
  -\Bigl( g(0)D^{(\mu)}_x f(x) + f(0) D^{(\mu)}_x g(x) \Bigr)-
 \]\vspace*{-5mm}
 \[ - \int\limits_0^1 {\rm d}t \left[ g'_t(x t^\mu)\Bigl(\int {\rm d}s f'_x(x s^\mu)
  \Bigr)_{s\to t} \,+ \right.
 \]
 \[\left.
 +\, f'_t(x t^\mu)\Bigl(\int {\rm d}s g'_x(x s^\mu)
  \Bigr)_{s\to t}\right]\!.
\]


\noindent Note that, for the functions with the property $g(0)=$
$=0$ and $f(0)=0$, this general formula somewhat simplifies.
  In the particular case of a product
  of monomials, the above formula (\ref{006}) is recovered.

\section{Different Ways\\ to Deform the Bose Gas Model}
As known, the system of standard oscillators (bosons) as an idealized model may be rather far from real physical systems.
Say, in case of a standard harmonic oscillator, the interaction is ignored.

Here, like many other papers studying deformed oscillators (e.g.,
\cite{ScarfoneSwamy08,AlginFibOsc,Monteiro}), we consider a
many-body system of deformed bosons or deformed oscillators. One of
the virtues of a deformation is the ability to effectively account for
the interaction between particles
\cite{ScarfoneSwamy08,ScarfoneSwamy09}, their non-zero volume
\cite{Avancini}, or their inner structure (composite nature)
\cite{GKM2011}. The joint account of two of these factors is also
possible \cite{GM2013}.

A special deformed Bose gas model that is the $\mu$-Bose gas model
associated with $\mu$-deformed oscillators \cite{Jann} was
introduced and studied in \cite{GavrRebIntercepts,GavrMishch}.
Therein, the main concern was the calculation of correlation
functions and their intercepts. In \cite{Arxiv}, the study of the
thermodynamics of a $\mu$-Bose gas \mbox{has begun.}

For the thermal average of an  operator $\mathcal{O},$ we use the
formula (that for the Hamiltonian is given below)
\begin{equation}\label{2}
\langle \mathcal{O} \rangle=\frac{Tr(\mathcal{O}e^{-\beta H})}{Z},
\end{equation}\vspace*{-3mm}

\noindent with $Z$ being the grand canonical partition function. Its
logarithm in the non-deformed case is
\begin{equation}\label{3}
\ln Z=-\sum_i\ln(1-ze^{-\beta\varepsilon_i})
\end{equation}\vspace*{-3mm}

\noindent with the fugacity $z=e^{\beta\widetilde{\mu}}$. The total
number of\, par-

\noindent ticles is calculated from the formula
\begin{equation*}\label{4}
N=z\frac{d}{dz}\ln Z,
\end{equation*}
which will be modified for the  purposes of the present treatment
(see the next section for the $\mu$-defor- \mbox{med case}).

For deformed oscillators and a $\mu$-deformed Bose gas,
the deformation parameter is inevitably present. Note that the building of a deformed analog
of the Bose gas model can be done in different ways:

--~By taking a modified/deformed Hamiltonian for the
system:
\begin{equation}
H=\sum_i(\varepsilon_i-\widetilde{\mu})[N_i],
\end{equation}

\noindent where $\varepsilon_i$ is the kinetic energy of particles,
$\widetilde{\mu}$ is the chemical potential, and $[X]\neq X$ stands
for the deformed bracket.

--~By proceeding with the usual Hamiltonian $H=$\linebreak
$=\sum_i(\varepsilon_i-\widetilde{\mu})N_i$ and then using the
(technique of) modified calculus.\,\,Namely, instead of the usual
derivative, one uses some its generalization to obtain
thermodynamical functions. Then all thermodynamical functions will
depend on the deformation \mbox{parameter.}

--~Here, we perceive another, specially designed way. Namely, we use the generalized
derivative only once: to obtain the expression for the total number of particle;
then we use this result to get the deformed partition function. Having derived the $\mu$-deformed partition function, we
 easily find all the other (also deformed) thermodynamical functions.

\section{\boldmath$\mu$-Deformation of the Bose Gas Model: Correlation Functions}
The system of standard oscillators or bosons, as an idealized model, usually is very distant from real physical systems.
As known, the standard harmonic oscillator implies the absence of interaction.

In our treatment, like many other papers studying the deformed
oscillators (see, e.g.,
\cite{Monteiro,ScarfoneSwamy09,AlginFibOsc}), we consider a set of
deformed oscillators or deformed bosons. Concerning the deformed
models, let us stress its ability to effectively account for  the
interaction between particles; their non-zero volume; and their
inner structure or composite nature.

The $\mu$-deformed analog of the Bose gas model ($\mu$-Bose gas) is
the model describing the many-particle system of $\mu$-deformed
bosons with the Hamiltonian\,\footnote[3]{We use the symbol
$\widetilde{\mu}$ for the chemical potential in order to distinguish
it from the deformation parameter $\mu$.}\vspace*{-1mm}
\begin{equation}\label{1}
H=\sum_i(\varepsilon_i-\widetilde{\mu})N_i,
\end{equation}\vspace*{-4mm}

\noindent which is similar to the case of ordinary bosons (the
simplest possible form). Here, $\varepsilon_i$ in (\ref{1}) denotes
the kinetic energy of a particle in the state ``$i$'', $N_i$ is the
particle number (occupation number) operator corresponding to the
state ``$i$''. The thermodynamic study of the new model will be
based on special mathematical techniques~-- the so-called
$\mu$-calculus \mbox{described above.}

The $\mu$-Bose gas model associated with $\mu$-deformed oscillators
\cite{Jann} was first introduced and studied in
\cite{GavrRebIntercepts}. This model has a potential for the
explanation of the non-Bose like behavior of the intercepts of the
momentum correlation functions of real particles such as
$\pi$-mesons. Some further development of the correlation function
(and intercept) aspects of the $\mu$-Bose gas model was made in
\cite{GavrMishch}. Note that, due to these efforts, the intercepts
of the second, third and higher order correlation functions are
known in explicit analytic form.\vspace*{-1mm}

\section{Thermodynamic Functions\\ of the \boldmath$\mu$-Bose Gas Model}
We now study the thermodynamics (see \cite{Arxiv}) of the
$\mu$-deformed analog of a Bose gas in more details, by using the
$\mu$-calculus given above. We consider non-relativistic particles.
Below, the regimes of high and low temperatures will be considered.
First of all, let us calculate the total number of
particles.\vspace*{-1mm}

\subsection*{Total number of particles}

As mentioned, the relation giving the total number of particles in
the Bose gas model is\vspace*{-2mm}
\begin{equation}\label{8}
N=z\frac{d}{dz}\ln Z.
\end{equation}\vspace*{-5mm}

\noindent To deal with the thermodynamics of a $\mu$-deformed analog
of the Bose gas model, formula (\ref{8}) for the total number $N$
should be modified. Namely, we take\vspace*{-1mm}
\begin{equation}\label{9-8}
N^{(\mu)}=z\mathcal{D}^{(\mu)}_z\ln
Z=-z\mathcal{D}^{(\mu)}_z\sum_i\ln(1-ze^{-\beta\varepsilon_i}),
\end{equation}\vspace*{-5mm}

\noindent where $\mathcal{D}^{(\mu)}$ is the $\mu$-derivative from
(\ref{7}).

We apply this, assuming $0\leq\mu<1$,
to the logarithm of the partition function in (\ref{3}) to get
\[N^{(\mu)}=z\sum_i\sum_{n=1}^{\infty}\frac{(e^{-\beta\varepsilon_i})^n}{n}[n]_{\mu}z^{n-1}=\]\vspace*{-5mm}
\begin{equation}\label{10-9}
=\sum_i\sum_{n=1}^{\infty}\frac{[n]_{\mu}}{n}(e^{-\beta\varepsilon_i})^nz^n.
\end{equation}
For consistency, we require $0\leq|ze^{-\beta\varepsilon_i}|<1$ in (\ref{10-9}).
Since non-relativistic particles are considered, the energy of the $i$-th particle of the system is
\begin{equation}\label{11-10}
\varepsilon_i=\frac{{\bf p}_i{\bf
p}_i}{2m}=\frac{|p|^2}{2m}=\frac{p_i^2}{2m},
\end{equation}
where ${\bf p}_i$ the 3-momentum of the particle in the $i$-th state, and
$m$ is particle's mass.

As seen, as $z\rightarrow 1,$ the expression in the summand in (\ref{10-9}) diverges
when $p_i=0, i=0$.
Below, we assume the $i=0$ ground state is associated with a macroscopically large occupation number.
Then, even though $z\neq 1$, we separate the term with $p_i=0$ from the remaining sum:
\begin{equation}\label{13-11}
N^{(\mu)}={\sum_i} ' \sum_{n=1}^{\infty}\frac{[n]_{\mu}}{n}(e^{-\beta\varepsilon_i})^nz^n +
\sum_{n=1}^{\infty}\frac{[n]_{\mu}}{n}z^n.
\end{equation}
The ``prime'' marking the outer sum in (\ref{13-11}) means that the $i=0$ term is dropped from the sum.
For large volumes $V$ and large $N$, the spectrum of single-particle states is almost a continuous one, and
we replace the sum in (\ref{10-9}) by the integral:
\begin{equation*}\label{12}
\sum_i\rightarrow \frac{V}{(2\pi \hbar)^3}\int d^3k.
\end{equation*}
In other words, we isolate the ground state, and
the contribution from all the other states is included in the integral.
WE now perform the integration over 3-momenta using spherical coordinates:
\[
  N^{(\mu)}=\frac{4\pi V}{(2\pi\hbar^2)^3}\sum_{n=1}^{\infty}\frac{[n]_{\mu}}{n}z^n \times
\]\vspace*{-5mm}
\begin{equation}\label{int}
\times  \int\limits^{\infty}_0 p^2e^{-\beta p^2/2m}dp
+\sum_{n=1}^{\infty}\frac{[n]_{\mu}}{n}z^n.
\end{equation}
The lower limit of the integral can still be taken as zero, because the ground state,
$p_0$, does not contribute to the integral anyway.
By performing the integration,
we finally obtain that the ($\mu$-deformed, i.e., depending on $\mu$)
total number of particles is given by the expression
\begin{equation}\label{14-12}
N^{(\mu)}\!=\!\frac{V}{\lambda^3}\sum_{n=1}^{\infty}\!\frac{[n]_{\mu}}{n^{5/2}}z^n\!+\!
N_0^{(\mu)}, ~~ N_0^{(\mu)}\!\equiv\!
\sum_{n=1}^{\infty}\!\frac{[n]_{\mu}}{n}z^n,
\end{equation}
which will be used for deriving the $\mu$-partition \mbox{function.}

\subsection*{Deformed partition function,\\ equation of state}

Let all the known relations between thermodynamical quantities known
in case of the usual Bose gas thermodynamics be promoted to its $\mu$-deformed
analog. That is, the well-known relations for
the usual Bose gas model and its $\mu$-deformed counterpart are formally similar. However,
all the thermodynamical quantities for the $\mu$-Bose gas model become $\mu$-dependent.

To obtain the $\mu$-partition function $\ln Z^{(\mu)}$ from
\begin{equation}\label{18-16+1}
N^{(\mu)}=z\frac{d}{dz}\ln Z^{(\mu)},
\end{equation}
we invert it:
\begin{equation}\label{19-17+1}
\ln Z^{(\mu)}=\bigg(\!z\frac{d}{dz}\!\bigg)^{\!-1}N^{(\mu)}.
\end{equation}
To perform the operation $\bigl(z\frac{d}{dz}\bigr)^{-1}$ and to get
$\ln Z^{(\mu)}$, we may either integrate $\ln
Z^{(\mu)}=\int\!dz~z^{-1}N^{(\mu)}$ or, equivalently, apply the
following property valid on the monomials $z^k$ for any function
$f(z\frac{d}{dz})$ possessing a power series expansion:
\begin{equation}\label{20-18+1}
f\bigg(\!z\frac{d}{dz}\!\bigg)z^k=f(k)z^k.
\end{equation}
In view of this, relations (\ref{19-17+1}) and
(\ref{13-11})--(\ref{14-12}) yield
\[
\ln
Z^{(\mu)}=\bigg(\!z\frac{d}{dz}\!\bigg)^{\!-1}\!\left(\!\frac{V}{\lambda^3}\sum_{n=1}^{\infty}\frac{[n]_{\mu}}{n^{5/2}}z^n+
\sum_{n=1}^{\infty}\frac{[n]_{\mu}}{n}z^n\!\right)\!=
\]\vspace*{-5mm}
\[
=\frac{V}{\lambda^3}\sum_{n=1}^{\infty}\frac{[n]_{\mu}}{n^{5/2}}\bigg(\!z\frac{d}{dz}
\bigg)^{\!-1}z^n+\sum_{n=1}^{\infty}\frac{[n]_{\mu}}{n}\bigg(\!z\frac{d}{dz}\bigg)^{\!-1}z^n=
\]\vspace*{-5mm}
\begin{equation}\label{21-19+1}
=\frac{V}{\lambda^3}\sum_{n=1}^{\infty}\frac{[n]_{\mu}}{n^{5/2}}(n)^{-1}z^n+\sum_{n=1}^{\infty}\frac{[n]_{\mu}}{n}(n)^{-1}z^n.
\end{equation}
The latter result can be written as (see (\ref{16-14}))
\begin{equation}\label{22-20+1}
\ln Z^{(\mu)}=\frac{V}{\lambda^3}g^{(\mu)}_{5/2}+g^{(\mu)}_1
\end{equation}
or, equivalently, as
\begin{equation}\label{22-20+2}
Z^{(\mu)}(z,T,V)=\exp\left(\!\frac{V}{\lambda^3}g^{(\mu)}_{5/2}(z)+g^{(\mu)}_1(z)\!\right)\!.
\end{equation}
Formulas (\ref{21-19+1})--(\ref{22-20+2}) for the $\mu$-deformed
partition function constitute our main result. Using (\ref{22-20+2}),
it is now possible to derive other thermodynamical functions.
The equation of state reads
\begin{equation}\label{23-21+1}
\frac{PV}{kT}=\ln Z^{(\mu)}=\frac{V}{\lambda^3}g^{(\mu)}_{5/2}(z)+g^{(\mu)}_1(z).
\end{equation}

Remark that an alternative way of obtaining the $\mu$-deformed partition function
could be the use of the one-particle $\mu$-deformed distribution function $\langle n\rangle^{(\mu)}$
found in analytical form
in \cite{GavrMishch}. Note that the result derived in such way will slightly
differ from that given in (\ref{22-20+2}). The details will be given elsewhere.

\subsection*{Virial expansion and virial coefficients}

Let us consider the physical meaning of the parameter $\mu$.
In the case of ideal Bose gas, the virial coefficients are responsible for the effective (two-particle, three-particle, {\it etc}.)
interactions of the quantum-correlation or quantum-statistical origin. In a $\mu$-Bose gas,
the inner structure/compositeness
of particles or interactions between them are effectively taken into account (like other deformations of the Bose
gas model) by means of the involved parameter $\mu$. Clearly,
this results in an additional effective interaction (revealed by the deformed virial coefficients).

Consider the regime of high temperature and low density $\lambda^3/v\ll 1$.
As explained, e.g., in \cite{Huang}, the second term in (\ref{14-12}) in this case is
negligibly small. This is also true for Eq.~(\ref{23-21+1}). Expression (\ref{14-12}) and the equation of state (\ref{23-21+1}) take the form
\begin{equation}\label{20-18}
\frac{\lambda^3}{v}=g^{(\mu)}_{3/2}(z), \quad \frac{Pv}{kT}=\frac{v}{\lambda^3}g^{(\mu)}_{5/2}(z),
\end{equation}
respectively. From the first relation in (\ref{20-18}), we have
\begin{equation}\label{21-19}
\frac{\lambda^3}{v}=z+\frac{[2]_{\mu}}{2^{5/2}}z^2+\frac{[3]_{\mu}}{3^{5/2}}z^3+
\frac{[4]_{\mu}}{4^{5/2}}z^4+\frac{[5]_{\mu}}{5^{5/2}}z^5+...,
\end{equation}
$[k]_{\mu}$ being the $\mu$-number corresponding to $k$.
Inverting the series in (\ref{21-19}), we derive the virial expansion for the equation of state
\begin{equation}\label{22-20}
\frac{Pv}{kT}=1+A\bigg(\!\frac{\lambda^3}{v}\!\bigg)+B\bigg(\!\frac{\lambda^3}{v}\bigg)^{\!2}+
C\bigg(\!\frac{\lambda^3}{v}\!\bigg)^{\!3}+D\bigg(\!\frac{\lambda^3}{v}\!\bigg)^{\!4}\!+...,
\end{equation}
where the second to fifth virial coefficients are:
\[
A=-\frac{[2]_{\mu}}{2^{7/2}[1]^2_{\mu}}, \quad B=\frac{[2]_{\mu}^2}{2^{5}[1]_{\mu}^4}-\frac{2[3]_{\mu}}{3^{7/2}[1]_{\mu}^3},
\]\vspace*{-5mm}
\[
C=-\frac{5[2]^3_{\mu}}{2^{17/2}[1]^6_{\mu}}+\frac{[2]_{\mu}[3]_{\mu}}{2^{5/2}3^{3/2}[1]^5_{\mu}}-\frac{3[4]_{\mu}}{2^{7}[1]^4_{\mu}},
\]\vspace*{-5mm}
\begin{equation}\label{D}
D\!=\!\frac{7[2]^4_{\mu}}{2^{10}[1]^8_{\mu}}-\frac{[2]^2_{\mu}[3]_{\mu}}{2^33^{3/2}[1]^7_{\mu}}+\frac{2[3]^2_{\mu}}{3^{5}[1]^6_{\mu}}+
\frac{[2]_{\mu}[4]_{\mu}}{2^{11/2}[1]^6_{\mu}}-\frac{4[5]_{\mu}}{5^{7/2}[1]^5_{\mu}}.
\end{equation}
The deformation parameter $\mu$ appears in the expressions for
virial coefficients in a specific manner, only through the
$\mu$-bracket of integers. We encounter a very unusual feature:
the appearance of the (powers of) $\mu$-unity $[1]_{\mu}$. Note that
its analog was absent in the virial coefficients for the cases of
a $q$-Bose or $p,q$-Bose gas because of the equality $[1]_{p,q}=1$,
see \cite{GR-12}. In the case at hands due to
$[1]_{\mu}=\frac{1}{1+\mu}\neq 1,$ the $\mu$-deformed virial
coefficients contain $[1]_{\mu}$ squared and also higher powers of
$\mu$-unity.

{{\bf Remark}}. By changing the value of deformation parameter, one
can regulate not only the magnitude, but also the sign of virial
coefficient(s) and, thus, even can get a repulsive instead of
attractive (or {\it vice versa}) effective interparticle interaction
in a model system. Generally speaking, by driving the value of
$\mu,$ we can effectively change (control) the very quantum
statistics of particles, compare with
\cite{ScarfoneSwamy09,AlginSenay,Ubriaco}.

\subsection*{Internal energy, specific heat,\\ and entropy at
\boldmath$T>T_c^{(\mu)}$}

For a deformed analog of the Bose gas model we adopt the known definition of
thermodynamical functions \cite{Pathria}. The internal energy
$U^{(\mu)}$ of a $\mu$-Bose gas is then found as
$U^{(\mu)}=-\left(\!\partial\ln
Z^{(\mu)}/\partial\beta\!\right)_{z,V}$. First, we examine the
case $T>T_c^{(\mu)}$:
\begin{equation}\label{002}
\frac{U^{(\mu)}}{N^{(\mu)}}=\frac{3}{2}\frac{kTv}{\lambda^3} g^{(\mu)}_{5/2}(z), \quad T>T_c^{(\mu)}.
\end{equation}
From the expressions for the internal energy of a $\mu$-Bose gas, we
obtain the {\it specific heat} of the system, using the relation
$C_v=\left(\partial U/\partial T \right)_{N,V}$. The result is
\begin{equation}\label{0002}
\frac{C_v^{(\mu)}}{N^{(\mu)}k}\!=\!\frac{15}{4}\frac{v}{\lambda^3} g^{(\mu)}_{5/2}(z)-\frac{9}{4}\frac{g^{(\mu)}_{3/2}{(z)}}{g^{(\mu)}_{1/2}{(z)}}, \quad T>T_c^{(\mu)}.
\end{equation}
Note that, both in (\ref{002}) and (\ref{0002}), the $\mu$-deformed analog (or $\mu$-polylogarithm) of the Bose function $g_n(z)$
does appear. This fact may have interesting implications.

The formula for the entropy, $S=\ln Z+\beta U-\beta\widetilde{\mu}
N,$ for the $\mu$-analog of a Bose gas in the regime of high
temperature reads
\begin{equation}\label{eh}
    \frac{S^{(\mu)}}{N^{(\mu)}k}=\frac{5}{2}\frac{V}{\lambda^3}g^{(\mu)}_{5/2}(z)
    - \ln z,
    \quad T>T^{(\mu)}_c.
\end{equation}
It will be interesting to explore the obtained formulas in different related contexts, and this is postponed for a subsequent report.

\subsection*{Critical temperature of condensation\\ and other
thermodynamical functions}

In the regime of low temperature and high density, let us obtain (say, like in \cite{GR-12} for the case of the $p,q$-Bose gas model) the
critical temperature $T^{(\mu)}_c$ of
the considered $\mu$-analog of the Bose gas model. We start with Eq.~(\ref{14-12}) and rewrite it as
\begin{equation}\label{01}
\frac{N_0^{(\mu)}}{V}=\frac{\lambda^3}{v}-g^{(\mu)}_{3/2}(z).
\end{equation}

\begin{figure}
\includegraphics[width=7cm]{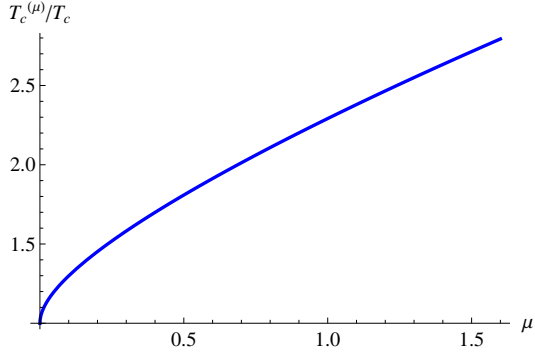}
\vskip-3mm\caption{Dependence of the ratio $T_c^{(\mu)}/{T_c}$ on
the $\mu$-parameter \label{Fig:3}}
\end{figure}

\noindent The critical temperature $T_c^{(\mu)}$ of a $\mu$-Bose gas is
determined from the equation $\lambda^3/{v}=g^{(\mu)}_{3/2}(1)$:
\begin{equation}\label{23-21}
T_c^{(\mu)}=\frac{2\pi\hbar^2/mk}{\bigl(vg^{(\mu)}_{3/2}(1)\bigr)^{2/3}}.
\end{equation}
From this, we infer the ratio of the critical temperature
$T_c^{(\mu)}$ to the usual Bose gas critical temperature~$T_c$:
\begin{equation}\label{24-22}
\frac{T_c^{(\mu)}}{T_c}=\Biggl(\frac{2.61}{g^{(\mu)}_{3/2}(1)}\!\Biggr)^{\!\!2/3}\!.
\end{equation}
Figure \ref{Fig:3} shows how the obtained ratio (\ref{24-22})
depends on the parameter $\mu$ (i.e., on the deformation strength).

Observe that, similarly to the case of a $p,q$-analog of the Bose gas model (see \cite{GR-12}),
the ratio $\frac{T_c^{(\mu)}}{T_c}$ has the important characteristic feature: the greater the deformation strength (here
given by $\mu$),
the higher is $T_c^{(\mu)}$. In the no-deformation limit $\mu\rightarrow 0,$ we have $T_c^{(\mu)}/{T_c}=1$.
That is, as $\mu\rightarrow 0$, the $\mu$-critical temperature goes over into the
usual one, $T_c^{(\mu)}\rightarrow T_c$ (a kind of consistency). If we drive
the deformation strength, we could raise the critical temperature significantly.

\vskip1mm

\begin{figure}
\vskip3mm
\includegraphics[width=7cm]{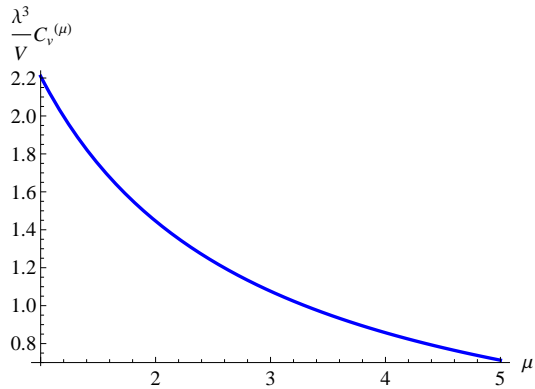}
\vskip-3mm\caption{Dependence of the specific heat $C_v^{(\mu)}$
(multiplied by $\frac{\lambda^3}{V}$) on the deformation parameter
$\mu$ \label{Fig-C}}\vskip2mm
\end{figure}

Now, we present the expressions for some thermodynamical functions, again for low temperatures. It is easy to
obtain the internal energy
\begin{equation}\label{U}
\frac{U^{(\mu)}}{N^{(\mu)}}= \frac{3}{2}\frac{kTv}{\lambda^3}
g^{(\mu)}_{5/2}(1), \quad T\leq T_c^{(\mu)},
\end{equation}
the specific heat
\begin{equation}\label{C}
\frac{C_v^{(\mu)}}{N^{(\mu)}k}= \frac{15}{4}\frac{v}{\lambda^3}
g^{(\mu)}_{5/2}(1), \quad T\leq T_c^{(\mu)},
\end{equation}
and the expression for the entropy (at $T\leq T_c^{(\mu)}$):
\begin{equation}\label{S}
   \frac{S^{(\mu)}}{N^{(\mu)}k}=\frac{5}{2}\frac{V}{\lambda^3}g^{(\mu)}_{5/2}(1),
    \quad T\leq T^{(\mu)}_c.
\end{equation}
In Fig. \ref{Fig-C}, the dependence of the specific heat on the parameter $\mu$ (deformation strength) is shown.

\section{Discussion, Outlook}\vspace*{-1mm}

For the $\mu$-analog of the Bose gas model proposed and studied earlier to some extent,
we have explored a number of main thermodynamical functions or relations. After presenting and discussing the necessary
elements of the $\mu$-calculus based on the use of the $\mu$-deformed analog of the usual derivative, we have
explicitly derived the expression for the total mean number of particles.
This important result allowed us to obtain the $\mu$-deformed partition function as well.
As a result, we have got via the
$\mu$-derivative that the deformation parameter has entered
the expressions for the total number of particles and the
partition function. While the $\mu$-deformed analog $g^{(\mu)}_{3/2}(z)$ of the polylogarithms $g  _{n}(z)$ becomes
involved in the formula for $N^{(\mu)}$, the $\mu$-polylogarithm
$g^{(\mu)}_{5/2}(z)$ has naturally appeared in the $\mu$-partition function.

Analyzing the high-temperature regime, we have obtained explicitly
the virial expansion for the equation of state along with the first
five virial coefficients. The deformation parameter $\mu$ enters the
expressions for virial coefficients through the $\mu$-bracket of
integers. Since the deformed system at $\mu\neq 0$ basically differs
from the bosonic one, it is obvious that a small variation of $\mu$
changes smoothly the statistics of particles. Moreover, by governing
the value of $\mu$, we are able to achieve such drastic change as
switching of the sign of virial coefficient(s), that implies the basic
change of the type of statistics. As it is obligatory for
consistency, one can easily verify that, in the no-deformation limit
$\mu=0,$ the known virial coefficients of the usual Bose gas  are
reproduced from the just obtained $\mu$-deformed virial
\mbox{coefficients.}

In the low temperature regime, the critical temperature (as a function
of $\mu$) of the analog of Bose condensation is obtained. The
dependence of the ratio $T_c^{(\mu)}/{T_c}$ on the deformation parameter
$\mu$ shows that the critical temperature of a $\mu$-Bose gas is
higher than $T_c$ of the usual Bose gas. This fact can be
obviously viewed as interesting and very useful for the goals of the
future more elaborated investigation of real Bose-like gases, in
parallel to a similar use of the result for the critical temperature
$T_c^{(p,q)}$ of the $p,q$-Bose gas model. Elsewhere, we hope to draw
some interesting consequences of our formulas for
\mbox{$C^{(\mu)}_{v}$ and $S^{(\mu)}$.}

\vskip2mm
 {\it This work was partly supported by the Special Program of the Division of Physics and Astronomy
of the NAS of Ukraine and (A.P.R.) by the Grant for Young Scientists of
the NAS of Ukraine (No.~0113U004910).}


\vspace*{-5mm}
  \rezume{А.П.\,Ребеш, І.І.\,Качурик, О.М.\,Гаврилик}
{ЕЛЕМЕНТИ $\mu$-ЧИСЛЕННЯ\\ ТА ТЕРМОДИНАМІКА МОДЕЛІ $\mu$-БОЗЕ-ГАЗУ} {На основі
$\mu$-деформованих осциляторів розроблено деформований аналог моделі
бозе-газу ($\mu$-бозе-газ). В рамках нової моделі в режимі низьких
температур ми  отримали віріальний розклад рівняння стану, а також
перші п'ять віріальних коефіцієнтів; в режимі низьких температур
обчислено критичну температуру конденсації. Ми також отримали питому
теплоємність, внутрішню енергію та ентропію для $\mu$-бозе-газу у
випадках низьких та високих температур. Усі термодинамічні функції
виявилися залежними від параметра деформації. Досліджено залежність
питомої теплоємності від параметра деформації.}

\end{document}